\documentclass[showpacs,onecolumn]{revtex4}

\usepackage{graphicx}                           % Permite incluir figuras e graficos no formato *.eps
\usepackage[ansinew]{inputenc}                  % Permite voce acentuar normalmente: í , ç , ã , ...
\usepackage{amsfonts}                           % Pacote para por simbolos matematicos especiais.
\usepackage[normalem]{ulem}                     % Pacote que permite sublinhar e/ou riscar palavras ou frases
\usepackage{amsmath}                            % Pacote que permite usar comandos especiaís nas fórmulas (como \dfrac).
\newcommand{\pla}{{\it Phys. Lett. A}}

\begin{document}

\author{J. G. G. de Oliveira Junior$^{1,2}$\footnote{Electronic address: zgeraldo@ufrb.edu.br}, J. G. Peixoto de Faria$^{3}$\footnote{
Electronic address: jgpfaria@des.cefetmg.br},
M. C. Nemes$^{2}$\footnote{Electronic address: carolina@fisica.ufmg.br}}%\email{carolina@fisica.ufmg.br}

\affiliation{$^1$Centro de Forma\c c\~ao de Professores,
Universidade Federal do
Rec\^oncavo da Bahia, 45.300-000, Amargosa, BA, Brazil\\
$^2$Departamento de F\'{\i}sica - CP 702 - Universidade
Federal de Minas Gerais - 30123-970 - Belo Horizonte - MG - Brazil\\
$^3$Departamento Acad\^emico de Disciplinas B\'asicas - Centro
Federal de Educa\c c\~ao Tecnol\'ogica de Minas Gerais - 30510-000 -
Belo Horizonte - MG - Brazil}

\title{Residual Entanglement and Sudden Death: a Direct Connection}


\begin{abstract}
%We show that for a particular class of initial states of tripartite
%systems, $|AB\rangle|C\rangle$ where $|AB\rangle$ is as entangled
%state, with unitary interaction between $A$ and $C$ there exists
%residual entanglement if sudden death of entanglement occurs in some
%partition.
We explore the results of Coffman {\it et.al.} [{\it Phys. Rev. A},
{\bf 61}, 052306 (2000)] derived for general tripartite states in a
dynamical context. We study a class of physically motivated
tripartite systems. We show that whenever entanglement sudden death
occurs in one of the partitions residual entanglement will appear.
For fourpartite systems however, the appearance of residual
entanglement is not conditioned by sudden death of entanglement. We
can only say that if sudden death of entanglement occurs in some
partition there will certainly be residual entanglement.
\end{abstract}

%\date{\today }

\maketitle

Entanglement, a property at the heart of Quantum Mechanics, has
first been brought to scientific debate the intriguing questions
posed by Einstein, Podolsky, and Rosen in ref. \cite{epr} and since
then the matter has always been under investigation. Recently the
interest of the physical community in this counterintuitive property
has raised even more due to its potential as a resource for
information processing and quantum computation \cite{nielsen}. For
that purpose having a profound knowledge of entanglement is a must
(see, {\it e.g.}, ref. \cite{emaranhamento} and references therein),
as well as a thorough comprehension of entanglement distribution in
composite systems (involving more than two degrees of freedom). In
this context, several years ago Coffman {\it et. al.}
\cite{wootters} studied the entanglement distribution in three qubit
systems ($ABC$), where each one of them can be entangled with the
other two. Moreover they proved the existence of what that they
called residual entanglement, which is not detected by usual two
qubits entanglement quantifiers \cite{wootters1,wootters2}. Their
result is valid for pure states in a Hilbert space
$2\otimes2\otimes2$, where they proved that quantum correlation
between $A$ and $BC$ will be manifest in one of three forms: $i$)
$A$ is entangled with $B$; $ii$) $A$ with $C$; and $iii$) the
entanglement is distributed among $ABC$, the so--called residual
entanglement. To this day this relation is the most general
available in the field of quantum information. In spite of its
mathematical rigor their relation has not yet been explored in
dynamical situations. We know that entanglement distribution is very
important for the implementation of quantum communication in
general, where the relevance of entanglement distribution is
crucial. This work is devoted to the purpose of understanding as
deeply as possible relevant dynamical consequences of the relation
derived in \cite{wootters}.

As the work of ref. \cite{wootters} was developed, an apparently
disconnected effect about entanglement has been found by
$\dot{\mbox{Z}}$yczkowski {\it et. al.} \cite{horodecki}. They very
recently showed that two parties entanglement can suddenly
disappear. Since then this dynamical characteristic of entanglement
has been called sudden death of entanglement
\cite{horodecki,morte,science_eberly} (hereafter ESD) and has been
measured \cite{medindo_morte} using twin photons. A step forward in
the solution to this question was given in an example studied by
Sainz {\it et. al.} in ref. \cite{distribuicao}. They studied a four
qubit system which interacts locally and pairwise and showed the
existence of an entanglement invariant. However in such systems
entanglement sudden death is also present (noted first in ref.
\cite{morte1}). What happens to the entanglement in a unitary
evolution in such a situation? Their result may point to the idea
that the amount of quantum correlations present in a closed system
should be conserved, however there is nothing to prevent a dynamical
redistribution of the initial entanglement. In others words, the
initial entanglement might migrate from one partition to others in a
way that the initial entanglement be conserved.

The purpose of the present work is to answer the following question:
what happens with the entanglement distribution when 3--qubit
systems undergo ESD? We show that residual entanglement is
intimately related to ESD for a large class of states.

Fourpartite systems are also investigated having the same question
in mind but no solid mathematical results to back up our model
result about the connection between ESD and appearance of genuine
entanglement. In this case environmental effects are taken in to
account.

The context of quantum optics the kind of interaction we use and our
modeling of reservoir effects has proven very realistic in many
situations of physical interest. Since entanglement dynamics is an
essential part of the implementation of quantum communication we
believe the results presented here may be of use.

\section*{{\it Tripartite--systems: Entanglement vs. Sudden Death}}

Let us consider a three qubit system $A$, $B$, $C$ where
entanglement can be found in all partitions. Coffman {\it et. al}
\cite{wootters} proved that quantum correlation between $A$ and $BC$
will be manifest as follows
\begin{equation}\label{tangle_A}
C_{A(BC)}^{2}=C_{AB}^{2}+C_{AC}^{2}+\tau_{ABC}
\end{equation}
where $\tau_{ABC}$ stands for a tripartite residual entanglement and
$C_{i(j)}$ is the concurrence between partitions $i$ and $j$.
Moreover the authors noticed that $\tau_{ABC}$ is invariant if one
interchanges $A$ and $B$. From the generality of eq.(\ref{tangle_A})
all entangled physical systems which may be mapped onto a three
qubits problem must obey (\ref{tangle_A}).
\begin{figure}[h]
\centering
  \includegraphics[scale=0.650,angle=00]{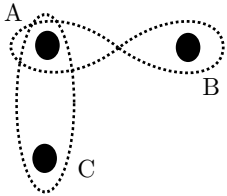}
  \caption{ System $ABC$ is initially in the pure state
  $|\psi_0\rangle=|AB\rangle|C\rangle$, with $AB$ initially
  entangled an factored of $C$. when a unitary interaction between $A$ and
  $C$ is turned on there will be an entanglement dynamics in the
  tripartite system.
}\label{fig}
\end{figure}

Initially for the physical situation depicted in fig. \ref{fig} we
have
\begin{equation}\label{cap_tangle_eq03}
C_{A(BC)}\,=\,C_{AB}\,=\,C_0
\end{equation}
and
\begin{equation}\label{cap_tangle_eq04}
 C_{AC}\,=\,\tau_{ABC}\,=\,0 \ \ \ .
\end{equation}

Let us consider now that the qubits $A$ and $C$ interact. When the
interaction is ``turned on"\,$A$ and $C$ will dynamically entangle
and, according to monogomy of entanglement \cite{monogamia}, $A$
will be less entangled with $B$. However, $C$ will ``see"\,the state
$A$ as
\begin{equation}\label{rhoA}
\rho_{A}=\mathrm{tr}_{B}\bigl(|AB\rangle\langle AB|\bigr) \ \ \ .
\end{equation}
As during this time evolution the partitions $AB$ and $C$ interact
one should expect that the entanglement distribution will be such
that $C_{AB}$ will become smaller as a function of time and  both
$C_{AC}$ and $C_{BC}$ start to grow accordingly. This is an example
where eq.(\ref{tangle_A}) must be obeyed all along the dynamics.
Therefore it is possible that besides $C_{AB}$, $C_{AC}$ and
$C_{BC}$ there may at some point appear a $\tau_{ABC}$. So, for
systems where the partition $AC$ admits interaction among its
constituents and shares $C_0$ with $B$ there are actually two very
enlightening dynamical situations: $i$) there is no ESD in any of
the partitions and $ii$) there is ESD in at least one of the
partitions. In the first case ($i$), with excitation exchange
between $A$ and $C$, one can show \cite{distribuicao1} that
\begin{equation*}
C_{AB}^{2}+C_{BC}^{2}=C_{0}^{2} \ \ \ .
\end{equation*}
Besides this result we also have that
\begin{equation}
C_{B(AC)}^{2}=C_{AB}^{2}+C_{BC}^{2}+\tau_{ABC}\label{tangle_B}
\end{equation}
must be obeyed so that
\begin{equation*}
C_{B(AC)}^{2}=C_{0}^{2}+\tau_{ABC} \ \ \ .
\end{equation*}
Now, since by hypothesis there is no ESD in any of the partitions
and $B$ does not interact with the partition $AC$, we have
$C_{B(AC)}^{}=C_{0}^{}$ and $\tau_{ABC}=0$. What happens when one of
the partitions undergoes ESD (ii)? In this case, given the
interaction between $A$ and $C$ the entanglement ($C_{AC}$) will not
disappear suddenly. Therefore ESD can only occur in partitions $AB$
and $BC$. Let us first consider that during a time interval ESD
occurs in partition $AB$. During this time window,
eq.(\ref{tangle_A}) gives
\begin{eqnarray*}
  C_{A(BC)}^{2}=C_{AC}^{2}+\tau_{ABC} \ \ \ .
\end{eqnarray*}
However, we should remark that $C$ ``sees" $A$ as a mixed state and
its capacity to entangle with $A$ will depend on how much $A$ is
entangled with $B$ and also on the type of interaction. Since
initially we have $C_{A(BC)}=C_0$ and knowing that $C_{A(BC)}^{}\geq
C_{AC}^{}$ during the whole evolution, in the interval when
$C_{AB}=0$ the residual entanglement $\tau_{ABC}$ must be different
from zero otherwise eq.(\ref{tangle_A}) will not be satisfied. The
same analysis is valid when ESD %sudden death of entanglement
occurs in the partition $BC$, from analyzing (\ref{tangle_B}). Last
but not least we consider the case in which ESD occurs in both
partitions. Then $C_{B(AC)}=C_0=\tau_{ABC}$.

The above considerations leave no doubt that the appearence of
entanglement sudden death \cite{horodecki,morte} in tripartite
systems bears very intimate connection with higher order
entanglement, {\it i.e.}, residual entanglement.

A concrete example is the tripartite system studied in ref.
\cite{2atomos1cavidade}, consisting of two atoms, only one of which
$A$, say, interacts with the cavity, the other $B$ serves the unique
purpose of allowing for an entangled initial state with $A$. The
cavity $C$ interacts resonantly with $A$ according to the usual
Jaynes--Cummings model \cite{jc}, where the interaction is given as
\begin{equation}\label{hjc}
H_{I}= \hbar g(c^{\dagger}\sigma_{-}^{A}+c\,\sigma_{+}^{A}) \ \ \ ,
\end{equation}
where $g$ is a coupling constant, $c$ ($c^{\dag}$) is an operator
that annihilates (creates) an excitation in $C$ and
${\sigma}_{-}^{A}=|\downarrow\rangle\langle \uparrow|$
(${\sigma}_{+}^{A}=|\uparrow\rangle\langle \downarrow|$) analogously
for the atoms. Consider the atomic initial state as given by
\begin{equation}\label{psi_AB}
|AB\rangle_{\psi}=\beta|\uparrow\downarrow\,\rangle+\alpha|\downarrow\uparrow\,\rangle
\end{equation}
and the cavity in vacuum $|C\rangle_0=|0\rangle$, with
$|\beta|^2+|\alpha|^2=1$. For this initial state
$|AB\rangle_{\psi}|C\rangle_0$ the evolved state will be
\begin{equation}\label{cap_tangle_eq09}
|ABC\rangle_{t}^{(0)}=\bigl[\beta\cos(gt)|\uparrow\downarrow\,\rangle+
\alpha|\downarrow\uparrow\,\rangle\bigr]|0\rangle-i\beta\sin(gt)|\downarrow\downarrow\rangle|1\rangle
\ \ \ .
\end{equation}
To quantify the entanglement between $A$ and $BC$ of state
(\ref{cap_tangle_eq09}), we will use concurrence in the form
$2\sqrt{\det\rho_{A}}\,$ \cite{wootters}, where
$\rho_{A}=\mathrm{tr}_{BC}(|ABC\rangle\langle ABC|_{t})\,$. For the
entanglement between $A$ and $B$ and between $A$ and $C$ we use the
concurrence which is defined in the refs. \cite{wootters1,wootters2}
as
\begin{equation}\label{cap01_eq9}
C_{\rho}=\max\bigl\{0,\sqrt{\lambda_1}-\sqrt{\lambda_2}-
\sqrt{\lambda_3}-\sqrt{\lambda_4}\bigr\}
\end{equation}
where the $\lambda_i$'s are the eigenvalues of the
$\rho\,(\sigma_{y}\otimes\sigma_{y})\,\rho^{*}\,(\sigma_{y}\otimes\sigma_{y})$
organized in decreasing order, $\sigma_{y}$ is one of the Pauli
matrices and $\rho^{*}$ is the complex conjugate of $\rho$. For the
state (\ref{cap_tangle_eq09}) each bipartition will have concurrence
\begin{eqnarray*}
  C_{AB}&=&C_{0}|\cos gt| \label{c13s}\\
  C_{AC}&=&|\beta|^2|\sin 2gt| \label{c23s}\\
  C_{BC}&=&C_{0}|\sin gt| \label{c33s}
\end{eqnarray*}
with $C_{0}=2|\beta\alpha|$ stands for the initial entanglement of
$AB$ and the entanglement between $A$ and $BC$ will be
\begin{eqnarray*}
C_{A(BC)}&=&2\sqrt{|\beta|^2\cos^2gt\,\bigl(|\alpha|^2+|\beta|^2
\sin^2gt\bigr)}  \ \ \ .
\end{eqnarray*}
For this initial state (\ref{cap_tangle_eq09}) there will be no ESD
in $AB$ and $AC$, since
\begin{eqnarray*}
 C_{AB}^{2}+C_{BC}^{2}=C_{0}^{2}
\end{eqnarray*}
and according to eq.(\ref{tangle_B})
\begin{eqnarray*}
  \tau_{ABC}=0 \ \ \ ,
\end{eqnarray*}
as discussed above. Otherwise we will also have
\begin{eqnarray}\label{ABCsemtangle}
C_{A(BC)}^{2}&=&C_{AB}^{2}+C_{AC}^{2}
\end{eqnarray}
showing explicitly that in this example $\tau_{ABC}=0$.

\begin{figure}[h]
\centering
  \includegraphics[scale=0.35,angle=-90]{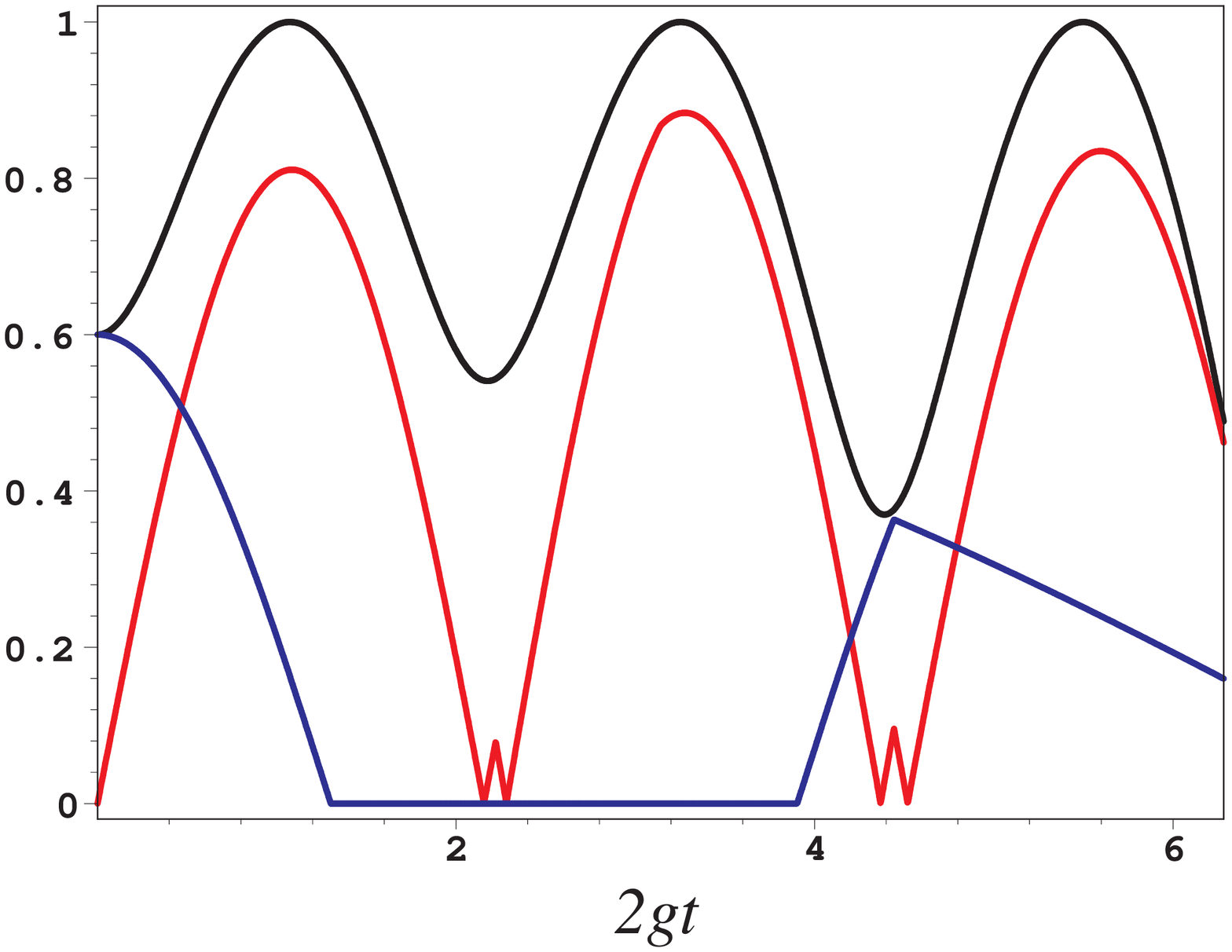}\hspace{20pt}
  \includegraphics[scale=0.35,angle=-90]{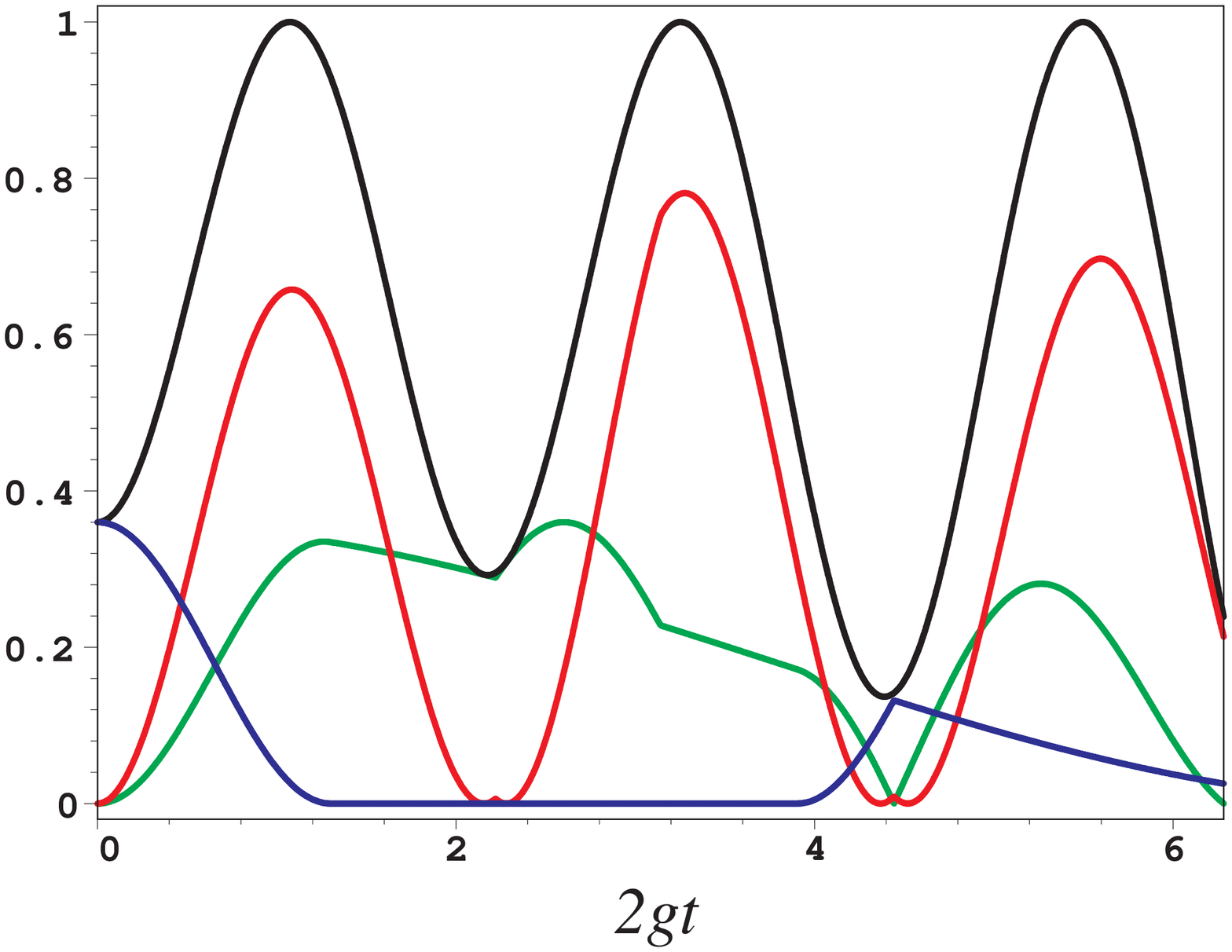}
  \caption{{\small Graphics for the state
  (\ref{cap_tangle_eq10}) with $\beta=3/\sqrt{10}$ and
  $\alpha=1/\sqrt{10}$. {\bf Left Figure (LF):} Concurrences in $AB$ and $AC$
  and between $A$ and $BC$. The blue, red and black curves are
  the concurrences $C_{AB}$, $C_{AC}$ and $C_{A(BC)}$, respectively.
  {\bf Right Figure (RF):} Here we show the residual entanglement and the concurrences
  squared between $A$ and $B$, $A$ and $C$, $A$ and $BC$. The green, blue, red and black curves are
  the residual entanglement $\tau_{ABC}$ and the concurrences squared $C_{AB}^{2}$,
  $C_{AC}^{2}$ and $C_{A(BC)}^{2}$, respectively.}}\label{comMSE}
\end{figure}

Next we consider an initial state which will dynamically be lead to
ESD in some partition. This will happen, {\it e.g.}, if the cavity
contains one excitation initially, $|C\rangle_1=|1\rangle$. For this
initial state the evolved state will be
\begin{equation}\label{cap_tangle_eq10}
|ABC\rangle_{t}^{(1)}=
\bigl[\beta\cos(\sqrt{2}gt)|\uparrow\downarrow\rangle
+\alpha\cos(gt)|\downarrow\uparrow\rangle\bigr]\,|1\rangle -
i\bigl[\alpha\sin(gt)|\uparrow\uparrow\rangle\,|0\rangle +
\beta\sin(\sqrt{2}gt)|\downarrow\downarrow\rangle\,|2\rangle\bigr]
\end{equation}
and the concurrences in $AB$ and $AC$ will be
\begin{eqnarray*}
  C_{AB} &=& C_{0}\max\bigl\{0,|\cos(gt)\cos(\sqrt{2}gt)| - |\sin(gt)\sin(\sqrt{2}gt)|\bigr\} \label{c13c}\\
  C_{AC} &=& \Bigl||\alpha|^2|\sin(2gt)|-|\beta|^2|\sin(2\sqrt{2}gt)|\Bigr|
  \label{c23c} \ \ \ .
%  C_{BC} &=&
%  C_0\Bigl||\sin(gt)\cos(\sqrt{2}gt)|-|\cos(gt)\sin(\sqrt{2}gt)|\Bigr|
%  \ \ \ ,
%  \label{c33c}
\end{eqnarray*}
It may be noted that ESD will be in the partition $AB$, as shown in
Figure \ref{comMSE}.
The entanglement between $A$ and $BC$ is
\begin{equation}\label{c43c}
C_{A(BC)}=2\sqrt{\Bigl(|\alpha|^2\sin^{2}(gt)+|\beta|^2\cos^{2}(\sqrt{2}gt)
\Bigr)
\Bigl(|\alpha|^2\cos^{2}(gt)+|\beta|^2\sin^{2}(\sqrt{2}gt)\Bigr)\,}
\end{equation}
and can not be written as (\ref{ABCsemtangle}). It is immediate that
$C_{A(BC)}^{2}\geq C_{AB}^{2}+C_{AC}^{2}$ and
\begin{equation}\label{tangle_morte}
\tau_{ABC}\geq 0 \ \ \ \ .
\end{equation}
This inequality reflects the main objective of this work. Observing
figure \ref{comMSE}, we note that the residual entanglement exists
right before of the ESD between $A$ and $B$. Our interpretation of
this result is as follows: the quantum correlations between $A$ and
$B$ disappear for a time interval and are distributed throughout the
system contributing to the residual entanglement \footnote{The same
reasoning follows if we considers as initial state
$|AB\rangle_{\phi}=\beta|\uparrow\uparrow\,\rangle+\alpha|\downarrow\downarrow\,\rangle$.}.

\section*{{\it Residual Entanglement and Sudden Death: a conjecture}}

\begin{figure}[h]
\centering
  \includegraphics[scale=0.750,angle=00]{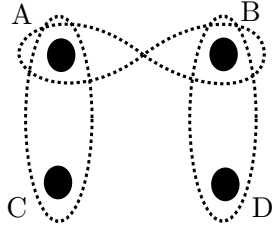}
  \caption{ Initially the partition $AB$ has $C_0$ of entanglement and is
  factored from $CD$. After an initial time a local interaction in the
  partitions $AC$ and $BD$ begins and there will be an entanglement
  dynamics in the system.}\label{fig2}
\end{figure}

We now consider a four partite system, $A$, $B$, $C$ and $D$ where
initial entanglement $C_0$ is in the partition $AB$. We also
consider that $A$ interacts locally with $C$, and $B$ with $D$ as
shown in figure \ref{fig2}. More concretely we consider two atoms
$A$ and $B$ sharing an entanglement $C_0$ and the partition $C$
consists of $N$ oscillators initially in vacuum, same for $D$. %This system is
%precisely the well studied double Jaynes--Cummings when $N=1$
%\cite{morte1,distribuicao} and in the limit $N\rightarrow\infty$ we
%have a reservoir with constant spectrum as studied in
%\cite{morte,medindo_morte,nascimento}.
%
The local interaction in the partition $AC$ will be described by the
hamiltonian
\begin{equation}\label{H_AC}
H_{AC}=\dfrac{\hbar\omega_A}{2}\sigma_{z}^{A}+\hbar\sum_{k=1}^{N}\omega_{k}
c^{\dag}_{k}c_{k}+\hbar\sum_{k=1}^{N}
g_k(c^{\dagger}_{k}\sigma_{-}^{A}+c_{k}\sigma_{+}^{A})
\end{equation}
where $g_{k}$ is a coupling constant between the atom and the
$k$--th oscillator of $C$, $c_{k}$ ($c^{\dag}_{k}$) is the operator
which annihilates (creates) one excitation in the $k$--th oscillator
of $C$ and ${\sigma}_{-}^{A}=|\downarrow\rangle\langle \uparrow|$
(${\sigma}_{+}^{A}=|\uparrow\rangle\langle \downarrow|$) in the
atom. Similarly for $BD$.

The fundamental state of the system $AC$,
$|\downarrow\,\rangle\prod_{k=1}^{N}|0_k\rangle$, does not evolve in
time. However the initial state containing one excitation,
$|\uparrow\,\rangle\prod_{k=1}^{N}|0_k\rangle$, evolves to the state
\begin{equation}\label{estado_10}
|\gamma(t)\rangle=\xi(t)|\uparrow\,\rangle|\tilde{0}\rangle
+\chi(t)|\downarrow\,\rangle|\tilde{1}\rangle
\end{equation}
where $\xi(t)$ and $\chi(t)$ are functions to be determined which
depend on $N$. We define the collective states
\begin{eqnarray}
  |\tilde{0}\rangle &=& \prod_{k=1}^{N}
|0_k\rangle \label{cap_tangle_eq11}\\
  |\tilde{1}\rangle&=&(1/\chi(t))\sum_{k=1}^{N}\lambda_k(t)|1_k\rangle
  \label{cap_tangle_eq12}
\end{eqnarray}
with  $|\chi(t)|^2=\sum_{k=1}^{N}|\lambda_k(t)|^2$ and
$|\xi(t)|^2+|\chi(t)|^2=1\,$. When $N=1$, in the resonant limit, we
have in $AC$ and $BD$ the so called Double Jaynes--Cummings
\cite{morte1,distribuicao}. The explicit forms for $\xi(t)$ and
$\chi(t)$ are
\begin{eqnarray}
  \xi(t)  &=& \cos(gt) \label{cap_tangle_eq13}\\
  \chi(t) &=& -i\sin(gt) \ \ \ . \label{cap_tangle_eq14}
\end{eqnarray}
Otherwise, when $N\longrightarrow\infty$ the subsystem $C$ is a
reservoir in vacuum and $A$ will decay exponentially as studied in
references \cite{morte,medindo_morte,nascimento}. In this case we
have
\begin{eqnarray}
  \xi(t)  &\longrightarrow& e^{-\gamma t/2} \label{cap_tangle_eq15}\\
  \chi(t) &\longrightarrow& \sqrt{1-e^{-\gamma t}\,} \label{cap_tangle_eq16}
\end{eqnarray}
where $\gamma$ is a damping constant.

Now let us consider the atoms prepared, as before, in $|AB\rangle_
{\psi}$ and the $2N$ oscillators in vacuum. This initial state
dynamically evolves to
\begin{equation}\label{psi_ABCD}
|ABDC\rangle_{t}^{(\psi)}=\beta|\gamma(t)\rangle_{AC}
|\downarrow\tilde{0}\rangle_{BD} + \alpha
|\downarrow\tilde{0}\rangle_{AC}|\gamma(t)\rangle_{BD} \ \ \ \ .
\end{equation}
The concurrencies of each pair are given by
\begin{eqnarray}
  C_{AB}&=&C_{0}|\xi(t)|^{2} \label{cap_tangle_eq17} \\
  C_{AC}&=&2|\beta|^{2}|\xi(t)\chi(t)| \label{cap_tangle_eq18}\\
  C_{AD}&=&C_{0}|\xi(t)\chi(t)| \label{cap_tangle_eq19}\\
  C_{BC}&=&C_{0}|\xi(t)\chi(t)| \label{cap_tangle_eq20}\\
  C_{BD}&=&2|\alpha|^{2}|\xi(t)\chi(t)| \label{cap_tangle_eq21}\\
  C_{CD}&=&C_{0}|\chi(t)|^{2} \ \ \ \label{cap_tangle_eq22}.
\end{eqnarray}
From eqs.(\ref{cap_tangle_eq17} -- \ref{cap_tangle_eq22}) we may
check that there will be no sudden death in any of the partitions of
$ABCD$. The concurrences between $A$ and the rest of the system is
given by
\begin{equation}\label{cap_tangle_eq23}
C_{A(BCD)}\,=\,2|\beta\xi(t)|\,\sqrt{|\alpha|^2+|\beta|^2|\chi(t)|^2\,}
\end{equation}
which can be rewritten as
\begin{equation}\label{C_A(BCD)}
C_{A(BCD)}\,=\,\sqrt{C_{AB}^{2}+C_{AC}^{2}+C_{AD}^{2}\,} \ \ \ ,
\end{equation}
{\it i.e.}, for the atoms initially prepared in the state
$|AB\rangle_{\psi}$  and the $2N$ oscillators in their vacuum state,
the entanglement that $A$ shares which the rest of the system is
completely distributed in the partitions $AB$, $AC$, and $AD$.
Therefore there will be no residual entanglement.

\begin{figure}[h]
\centering
  \includegraphics[scale=0.35,angle=-90]{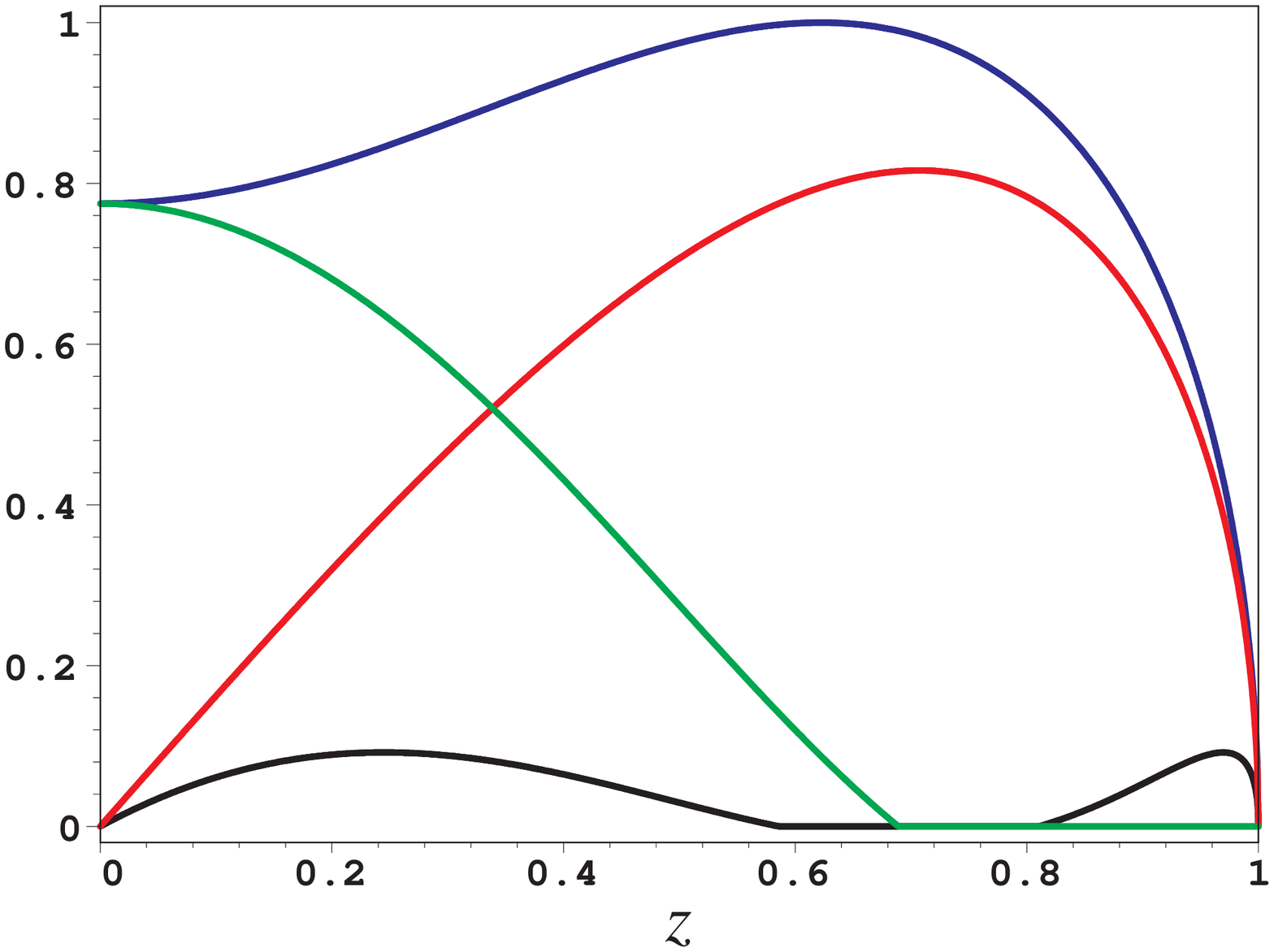}\hspace{20pt}
  \includegraphics[scale=0.35,angle=-90]{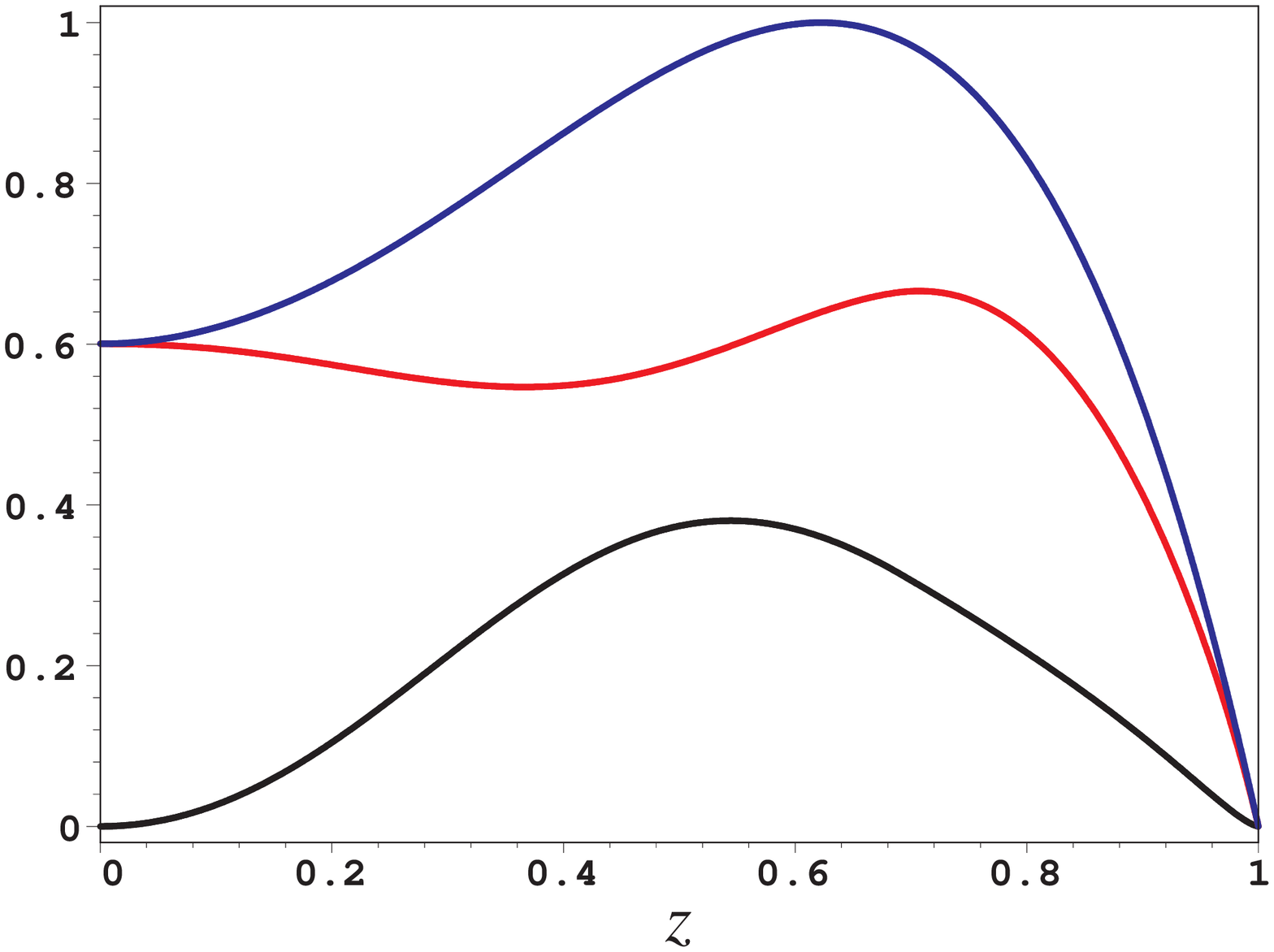}
  \caption{{\small Concurrences as a function of $z=|\chi(t)|$, with
  $\beta\approx0.905$ an $\alpha\approx0.429$ ($|\beta|>2|\alpha|$) in the state
  (\ref{phi_ABCD}). {\bf LF}: $C_{A(BCD)}$ in blue, $C_{AC}$ in red,
  $C_{AD}$ in black, and $C_{AB}$ in green. {\bf RF}: $C_{A(BCD)}^{2}$ in
  blue, $C_{AB}^{2}+C_{AC}^{2}+C_{AD}^{2}$ in red, and $E_{ABCD}$ in
  black.%
}}\label{ABCD1}
\end{figure}

A qualitatively different situation arises if one considers the
initial state
\begin{equation}\label{phi_AB}
|AB\rangle_{\phi}=\beta|\uparrow\uparrow\,\rangle+\alpha|\downarrow\downarrow\,\rangle
\end{equation}
for the atoms and the $2N$ oscillators in vacuum. This initial state
evolves to the state
\begin{equation}\label{phi_ABCD}
|ABDC\rangle_{t}^{(\phi)}=\beta|\gamma(t)\rangle_{AC}|\gamma(t)\rangle_{BD}+
\alpha
|\downarrow\tilde{0}\rangle_{AC}|\downarrow\tilde{0}\rangle_{BD} \ \
\ \ .
\end{equation}
It is well known that this initial condition, for $N=1$
\cite{morte1,distribuicao} and $N\rightarrow\infty$
\cite{morte,medindo_morte,nascimento}, presents ESD in some
partition when $|\beta|>2|\alpha|$. We focus our attention on the
entanglement that $A$ shares with the rest of the system. The
entanglement between $A$ and any other subsystem and that of $A$
with $BCD$ are given by
\begin{eqnarray}
  C_{AB}&=&2|\beta\xi(t)^2|\max\{0,|\alpha|-|\beta\chi(t)^2|\} \label{cap_tangle_eq24} \\
  C_{AD}&=&2|\beta\xi(t)\chi(t)|\max\{0,|\alpha|-|\beta\xi(t)\chi(t)|\} \label{cap_tangle_eq25}\\
  C_{AC}&=&2|\beta^2\xi(t)\chi(t)| \label{cap_tangle_eq26}\\
  C_{A(BCD)}&=&2|\beta\xi(t)|\sqrt{|\beta\chi(t)|^2+|\alpha|^2\,} \ \
  \ .\label{cap_tangle_eq27}
\end{eqnarray}
It becomes apparent that the entanglement in partitions $AB$ and
$AD$ may disappear suddenly. In the partition $AB$ there will be ESD
for times such that
\begin{equation}\label{cap_tangle_eq28}
|\chi(t)|^2\,\geq\,\biggl|\dfrac{\alpha}{\beta}\biggr| \ \ \ .
\end{equation}
In $AD$, ESD will occur at times such that
%$|\chi(t)|\sqrt{1-|\chi(t)|^2\,}\,=\,|\alpha/\beta|$ and $C_{AD}=0$
%when
%
\begin{equation}\label{cap_tangle_eq29}
|\chi(t)|\sqrt{1-|\chi(t)|^2\,}\,\geq\,\biggl|\dfrac{\alpha}{\beta}\biggr|
\end{equation}
where we used the fact that $|\xi(t)|=\sqrt{1-|\chi(t)|^2\,}$. When
we solve the inequality in (\ref{cap_tangle_eq29}) we find
\begin{equation}\label{cap_tangle_eq30}
\dfrac{1}{2}-\sqrt{\dfrac{1}{4}-\biggl|\dfrac{\alpha}{\beta}\biggr|^2}
<|\chi(t)|^2<\dfrac{1}{2}+\sqrt{\dfrac{1}{4}-\biggl|\dfrac{\alpha}{\beta}\biggr|^2}
\end{equation}
which imposes the condition  $|\beta|>2|\alpha|$. So when we observe
the inequalities  (\ref{cap_tangle_eq28}) and
(\ref{cap_tangle_eq30}) it is easy to see that when
\begin{equation}\label{cap_tangle_eq31}
\biggl|\dfrac{\alpha}{\beta}\biggr|\leq|\chi(t)|^2\leq\dfrac{1}{2}+\sqrt{\dfrac{1}{4}-\biggl|\dfrac{\alpha}{\beta}\biggr|^2}
\end{equation}
with $|\beta|>2|\alpha|$, $C_{AB}=0$ and $C_{AD}=0$ at the same
time. This will always be the case for the initial state
(\ref{phi_AB}) with $|\beta|>2|\alpha|$, as shown in figure
\ref{ABCD1}.

The entanglement between $A$ and $BCD$ may be rewritten as
\begin{equation}\label{C_A(BCD)}
C_{A(BCD)}^{2}\,=\,C_{AC}^{2}+C_{0}^{2}|\xi(t)|^2 \ \ \ ,
\end{equation}
which is valid during the whole evolution. The relationship above
shows that the entanglement shared by $A$ with the rest of the
system is divided in two parts. i) one which is in $AC$ due to the
local interactions between $AC$ and ii) one which is spread over the
rest of the system. We may now define the positive semidefinite
quantity \cite{osborne} $E_{ABCD}$ which represents the entanglement
between $A$ and $BCD$ which cannot be accounted for by the
entanglement of $A$ with $B$, $C$ and $D$ separately, {\it i.e.},
$E_{ABCD}=C_{A(BCD)}^{2}-\bigl[C_{AB}^{2}+C_{AC}^{2}+C_{AD}^{2}\bigr]$,
which for our case gives
\begin{equation}\label{cap_tangle_eq32}
E_{ABCD}=C_{0}^{2}|\xi(t)|^2-\bigl[C_{AB}^{2}+C_{AD}^{2}\bigr] \ \ \
.
\end{equation}

Fig. \ref{ABCD1} illustrates the entanglement distribution in the
case of the initial condition (\ref{phi_ABCD}). Note (on the LF)
that ESD only occurs in the partions $AB$ and $AD$. When $0.584
 \lesssim  z \lesssim 0.812$, $C_{AD}=0$. However when $0.689\lesssim z$ we
will have $C_{AB}=0$. In this situation when $0.689\lesssim
z\lesssim  0.812$ we will have $C_{AB}=C_{AD}=0$ at the same time.
The {\bf RF} illustrates the behavior of $C_{A(BCD)}^{2}$. It shows
a smooth behavior when it is increasing or decreasing. It represents
all the entanglement between $A$ and $BCD$ including the one coming
from the unitary interaction. The curve in red
$C_{AB}^{2}+C_{AC}^{2}+C_{AD}^{2}$ initially decreases due to the
entanglement decrease followed by ESD in $AB$ and $AD$. Right after
that it increases since the entanglement provided by the interaction
in the partition $AC$ becomes quantitatively significant. $E_{ABCD}$
(curve in black) presents a maximum before the other graphs. This is
due to the fact that the entanglement between $AB$ and between $AD$
are decreasing and the $AC$ entanglement is not yet qualitatively
significant. After the ESD in $AB$ and $AD$ the entanglement due to
the $AC$ dynamics grows, so that the $E_{ABCD}$ curves starts to
decrease.

Interestingly enough the $E_{ABCD}$ entanglement will be present
during the whole evolution for any value of $\beta$ and $\alpha$ in
state (\ref{phi_ABCD}). This means that for the initial state
$|AB\rangle_{\phi}$ there will always be an entanglement between $A$
and $BCD$ which cannot be accounted for by the entanglement of $A$
with $B$, $C$, $D$ separately. This is not true for the initial
state $|AB\rangle_{\psi}$ where we have $E_{ABCD}=0$ and all the
entanglement content between the partitions $A$ and $BCD$ may be
accounted for by two partite concurrences. When $|\beta|>2|\alpha|$
in the initial state (\ref{phi_AB}) there will be a time interval
$\Delta t$, definede by eq. (\ref{cap_tangle_eq31}), during which
$C_{AB}=0$ and $C_{AD}=0$, as discussed above. In this situation we
have
\begin{equation}\label{cap_tangle_eq33}
E_{ABCD}=C_{0}^{2}|\xi(t)|^2
\end{equation}
which represents the entanglement distributed in the whole system
that cannot be accounted for by $C_{AB}$, $C_{AC}$, and $C_{AD}$.

\section*{{\it Discussion and Conclusion}}

The interaction presented here where excitations are exchanged
between the atoms and the field, simulate quantum circuits in
Quantum Optics \cite{exploring}. This interaction provides for the
possibility of exchange information \cite{memoria} and also transfer
of entanglement in systems like those represented here \cite{swap}.
For example in eq. (\ref{H_AC}) in the limit when $N \longrightarrow
\infty $ with system $C$ initially in the vacuum state under the
well know Born--Markov approximation \cite{noise} simulates the
vacuum fluctuations responsible for the atomic exponential decay. In
spite of its broad range of applicability, this type of interaction
does not cover all phenomena in Quantum Optics. A phase coupling
between $A$ and the $N$ oscillators is also a useful kind of
environment without excitation exchange. When $N \longrightarrow
\infty $ this dynamics leads to the disappearance of coherence
\cite{carmichael} and may also induce ESD. In such situation one
should expect that the entanglement distribution be very similar to
the genuine entanglement in tripartite systems. This phase
interaction between atoms and fields (when $N\longrightarrow\infty$)
is similar to the one modeled in ref. \cite{oc_eberly} where ESD is
observed when two entangled atoms are subjected to a classical noisy
environment simulated by a stochastic classical field. This results
in phase damping of the collective and individual atomic states.

Recent studies show the existence of ESD in systems qubits--qutrits
($2\otimes 3$) \cite{pla1,found} and sudden death of nonlocality in
three qubit systems ($2\otimes2\otimes2$) \cite{found,pla2,pla3}
have also been investigated. In the first case, as also discussed
here, if the phase reservoir interacts either locally on globally
with the qubit and qutrit, then one should expect residual
entanglement in the subsystem of interest. The second case, where a
sudden disappearance of nonlocality is observed, requires, however a
more careful analysis since there exists entangled states which do
not violate Bell inequalities \cite{werner,barret,toner,daniel} even
when they are tripartite with residual entanglement \cite{toth}. In
other words the investigation between entanglement and nonlocality
is a very promising, open area of research.

As for the present work we have shown that tripartite systems
subjected to a local interaction will exhibit a very close
connection between ESD and residual entanglement, based on eq.
(\ref{tangle_A}).

A four partite system has also been investigated and the same
phenomenon is observed, Next a natural conjecture is in order: is
ESD a general mechanics through which entanglement flows from
partitions involving two qubits to larger ones? This is our belief
based on the fact that it can be rigorously demonstrated for three
qubits and several examples involving more qubits point in the same
direction. Proving this conjecture remain an open intriguing
challenge.

The authors JGPF and MCN were partially supported by the brazilian
agencies FAPEMIG (grant number CEX-PPM-00549-09) and CNPq (Instituto
Nacional de Ciência e Tecnologia em Informação Quântica).

\end{document}